\documentclass{emulateapj}

\usepackage{amssymb}
\usepackage{natbib}

\shorttitle{Gravitational instability in the dust layer}
\shortauthors{Yamoto \& Sekiya}

\begin{document}

\title{Two evolutional paths of an axisymmetric gravitational \\ instability in the dust layer of a protoplanetary disk}

\author{Fumiharu Yamoto\altaffilmark{1}}
\affil{Division of Theoretical Astronomy, National Astronomical Observatory of Japan, \\ Osawa 2-21-1, Mitaka, Tokyo 181-8588, Japan}

\email{yamoto@th.nao.ac.jp}

\and

\author{Minoru Sekiya}
\affil{Department of Earth and Planetary Sciences, Faculty of Sciences, \\ Kyushu University, Hakozaki, Higashi-ku, Fukuoka 812-8581, Japan}

\email{sekiya@geo.kyushu-u.ac.jp}

\altaffiltext{1}{Research Fellow of the Japan Society for the Promotion of Science}

\begin{abstract}
Nonlinear numerical simulations are performed to investigate the density evolution in the dust layer of a protoplanetary disk due to the gravitational instability and dust settling toward the midplane. We assume the region where the radial pressure gradient at equilibrium is negligible so that the shear-induced instability is avoided, and also restrict to an axisymmetric perturbation as a first step of nonlinear numerical simulations of the gravitational instability. We find that there are two different evolutional paths of the gravitational instability depending on the nondimensional gas friction time, which is defined as the product of the gas friction time and the Keplerian angular velocity. If the nondimensional gas friction time is equal to 0.01, the gravitational instability grows faster than dust settling. On the other hand, if the nondimensional gas friction time is equal to 0.1, dust aggregates settle sufficiently before the gravitational instability grows. In the latter case, an approximate analytical calculation reveals that dust settling is faster than the growth of the gravitational instability regardless of the dust density at the midplane. Thus, the dust layer becomes extremely thin and may reach a few tenth of the material density of the dust before the gravitational instability grows.
\end{abstract}


\section{INTRODUCTION}

One of unsolved issues in the planetary system formation is how kilometer-sized planetesimals form. There are two controversial models of the planetesimal formation: one is the growth through mutual sticking of dust aggregates due to nongravitational forces (\citealp{wc93}; \citealp{bw00}; \citealp{sir04}), and another is the fragmentation of the dust layer due to the gravitational instability (GI) (\citealp{saf69}; \citealp{gw73}; \citealp{sek83}).

Micron-sized dust particles in a protoplanetary disk grow due to the thermal motion in the initial stage of the dust evolution (\citealp{hn75}; \citealp{bet00}). Unless the protoplanetary disk is globally turbulent, dust aggregates sweep up dust particles during the settling toward the midplane and the radial drift toward the central star, and grow to centimeter-sized aggregates (\citealp{wei80}; \citealp{nsh86}; \citealp{nn06}). When the dust density, which is defined as the total mass of the dust per unit volume of the disk, at the midplane exceeds a few tenth of the gas density, a velocity shear in the vertical direction grows. The shear-induced instability will occur due to the velocity shear and prevents dust settling (\citealp{wei80}; \citealp{cdc93}). Thus, the dust density at the midplane is difficult to approach the critical density of the GI, which is several hundreds of the gas density.

A few possibilities to exceed the critical density of the GI have been considered: (1) If dust aggregates grow to the radii of the order of 10 $\mbox{m}$, the coupling of the dust and gas becomes weak and the dust density at the midplane exceeds the critical density of the GI (Cuzzi et al. 1993; \citealp{ddc99}). However, \citet{wei95} has noted that the GI would be prevented by radial random velocities of the dust. (2) If the dust-to-gas surface density ratio increases by an order of magnitude through the dust migration or gas dissipation, the shear-induced instability is suppressed (\citealp{sek98}; \citealp{ys02}; \citealp{yc04}). (3) If the pressure has a local maximum value in a protoplanetary disk, the radial pressure gradient which causes the shear-induced instability would be negligible \citep{hb03}. In this letter, we focus on the last case and other possibilities will be considered in future works.


We have investigated the density evolution in the dust layer due to the GI and dust settling. In \S2, a model of numerical simulations is described. In \S3, results of numerical simulations and an approximate analytical calculation are shown. Finally, these results are summarized in \S4.

\begin{figure*}[htbp]
  \begin{center}
   \includegraphics[width=6.5in]{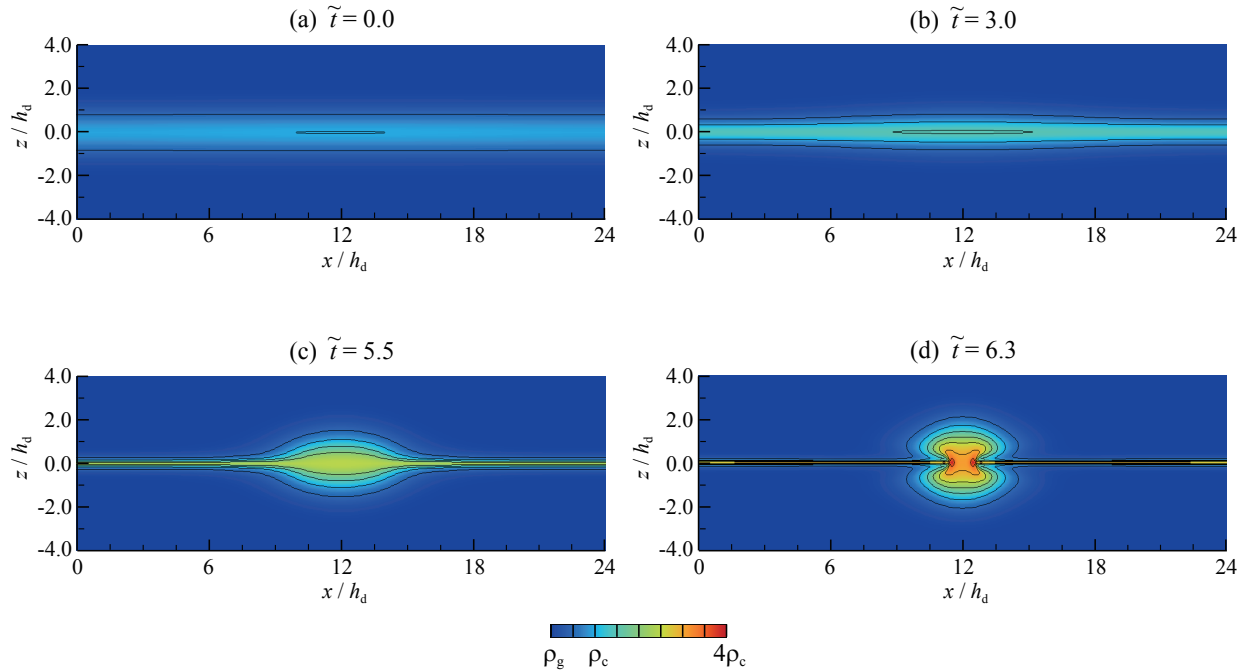}
  \end{center}
  \caption{The density evolution by the GI and dust settling for $\tilde{t}_{\rm{f}}=0.01$. Figures show the snapshots of the density at (a) $\tilde{t}=0.0$ (i.e., the initial condition), (b) $\tilde{t}=3.0$, (c) $\tilde{t}=5.5$, and (d) $\tilde{t}=6.3$. The length scales in the $x$ and $z$ directions are normalized by the scale height of the dust layer $h_{\rm{d}}$. The gas is incompressible and has a constant density. In this simulation, we assume $\rho_{\rm{c}}/\rho_{\rm{g}}=340$. This value corresponds to the gas density at 1 $\rm{AU}$ for the minimum mass solar nebula \citep{hay81}, for which we have $h_{\rm{d}}=86$ $\rm{km}$. The numerical simulation is performed for the region between the midplane ($z=0$) and the upper boundary ($z=4.0h_{\rm{d}}$), and lower halves of figures are drawn by using mirror symmetry with respect to the midplane.}
  \label{fig1}
  \vspace{0.2in}
\end{figure*}

\section{MODEL}

We use the local Cartesian coordinate system ($x$, $y$, $z$) rotating with $\Omega_{\rm{K}}$ at a representative distance $R$ from the central star, where $\Omega_{\rm{K}}$ is the circular Keplerian angular velocity, instead of the inertial system ($r$, $\phi$, $z$) with its origin at the central star. We use approximations in which only linear terms of ($x$, $y$, $z$) are left and higher order terms are neglected: $r \approx R+x$ and $y \approx R \phi$. In this letter, the dust layer is assumed to be axisymmetric with respect to the rotation axis, i.e., any physical quantity is independent of $y$. We assume the region where the radial pressure gradient at equilibrium is negligible so that the shear-induced instability is avoided.

The equations of continuity and motion of the dust and gas, and the Poisson equation for the self-gravitational potential are solved numerically. The tidal, the Coriolis and the self-gravity forces on the dust and gas are applied. Additionally, the gas drag force on the dust and its reaction and the pressure gradient forces on the gas are applied. The gas drag and its reaction forces depend on the gas friction time $t_{\rm{f}}$, which is defined as the time during which the relative velocity between a dust aggregate and gas becomes $1/e$. The dust pressure gradient force is assumed to be neglected because random velocities of dust aggregates are sufficiently suppressed by the gas drag force before the GI grows if $t_{\rm{f}} \ll \Omega_{\rm{K}}^{-1}$. All dust aggregates are assumed to have the same gas friction time which is constant for time. The gas is assumed to be incompressible. This assumption is good in the dust layer as explained in \citet{sek83}.


The Gaussian dust density distribution in the direction vertical to the midplane is used as the unperturbed density. The initial value of the density at the midplane is assumed to be equal to the critical density of the GI $\rho_{\rm{c}}$ \citep{ys04}. Unperturbed velocities of the dust and gas are set to zero in the $x$ direction and the Keplerian velocity in the $y$ direction because we assume that the radial pressure gradient is negligible. In the $z$ direction, the unperturbed gas velocity is set to zero and the unperturbed dust velocity is given by the terminal settling velocity:
\begin{equation}
  w_{\rm{d}} = - t_{\rm{f}} \Bigl\{\Omega_{\rm{K}}^2 z + 4 \pi G \int^{z}_{0}(\rho_{\rm{g}} + \rho_{\rm{d}})dz\Bigr\}, \label{wd}
\end{equation}
\\
\noindent
where $\rho_{\rm{g}}$ and $\rho_{\rm{d}}$ is the density of the gas and dust, respectively.

As an initial condition, we give only the dust density perturbation which has the critical wavelength of the GI obtained by the linear analysis \citep{ys04} and the maximum amplitude of a hundredth of the unperturbed dust density. The perturbed velocities of the dust and gas are set to be zero. In numerical simulations, boundaries in the $x$ direction are assumed to have a period of the critical wavelength. We assume that the gas density and the dust surface density are equal to those of the minimum mass solar nebula model (\citealp{hay81}; \citealp{hnn85}) at 1 $\rm{AU}$, but results are not sensitive to these assumptions and the only important parameter is the nondimensionalized gas friction time $\tilde{t}_{\rm{f}}$ (hereafter we denote nondimensional time and density by diacritic tilde, which are normalized by $\Omega_{\rm{K}}^{-1}$ and $\Omega_{\rm{K}}^2/(4 \pi G)$ respectively).


\section{RESULTS} 

The density evolution for $\tilde{t}_{\rm{f}}=0.01$ is shown in Figure \ref{fig1}. In the early stage of the numerical simulation (see Figs. \ref{fig1}a and \ref{fig1}b), the growth rate of the GI is so small that the density around the midplane mainly increases by dust settling. Next, the GI grows faster than dust settling and the dust layer in the calculated region changes into a ring circling the central star (see Fig. \ref{fig1}c). Afterwards, the density has a maximum value on the left and right in the ring and concave density distributions arise in the vicinity of a maximum density (see Fig. \ref{fig1}d). Subsequent evolution of the density is significant, but we should perform three dimensional numerical simulations because nonaxisymmetric modes may grow in this stage.

\begin{figure}[htbp]
  \begin{center}
   \includegraphics[width=3.25in]{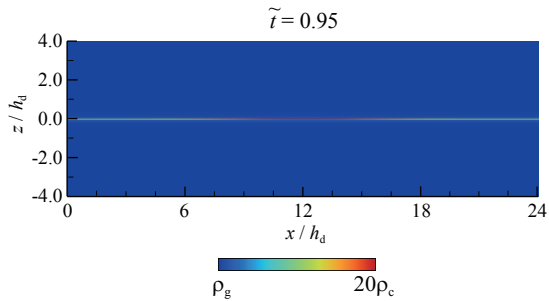}
  \end{center}
  \caption{The snapshot of the density at $\tilde{t}=0.95$ for $\tilde{t}_{\rm{f}}=0.1$. The length scales, the gas density, the upper boundary, and the simulated region are the same as Figure \ref{fig1}. Note that the maximum value of the density is different from Figure \ref{fig1}.}
  \label{fig2}
  \vspace{0.2in}
\end{figure}

The snapshot of the density at $\tilde{t}=0.9$ for $\tilde{t}_{\rm{f}}=0.1$ is shown in Figure \ref{fig2}. The settling velocity of dust aggregates (eq. [\ref{wd}]) for $\tilde{t}_{\rm{f}}=0.1$ is ten times as large as that for $\tilde{t}_{\rm{f}}=0.01$ due to the difference of the gas friction time. The numerical simulation shows that dust settling is faster than the growth of the GI as long as the dust density at the midplane is smaller than about twenty times of the critical density of the GI.

\begin{figure}[htbp]
  \begin{center}
   \includegraphics[width=3.25in]{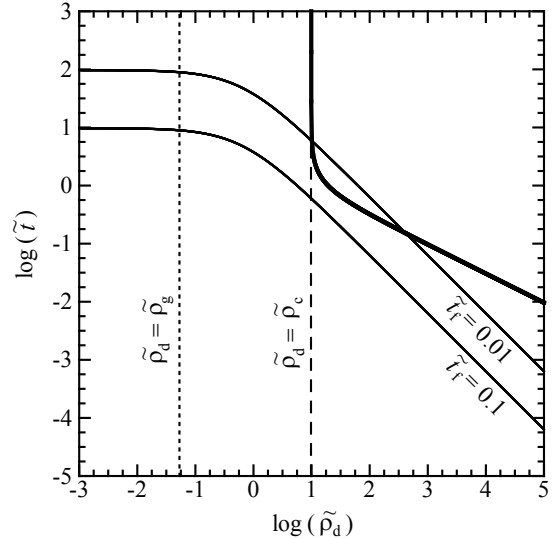}
  \end{center}
  \caption{Comparison between the dust settling time toward the midplane with $\tilde{t}_{\rm{f}} = 0.01$ (upper solid line) and $0.1$ (lower solid line) and the growth time of the GI with the Gaussian dust density distribution (bold line). The gas density $\tilde{\rho}_{\rm{g}}$ is shown by the dotted line. The critical density of the GI $\tilde{\rho}_{\rm{c}}$ is shown by the dashed line.}
  \label{fig3}
  \vspace{0.2in}
\end{figure}

For $\tilde{t}_{\rm{f}}=0.1$, dust settling is faster than the growth of the GI within the time interval of the numerical simulation. However, the numerical simulation is performed for the density only up to about twenty times the critical density. Here, we show that dust settling is faster than the growth of the GI regardless of the dust density at the midplane. The dust settling time $\tilde{t}_{\rm{settle}}$ during which the dust density increases by a factor of $e$ is given by

\begin{equation}
  \tilde{t}_{\rm{settle}} = \frac{1}{(1+\tilde{\rho}_{\rm{g}})\tilde{t}_{\rm{f}}}
    \Biggl\{1-\mbox{ln}\biggl(\frac{\tilde{\rho}_{\rm{d}}e+1+\tilde{\rho}_{\rm{g}}}
      {\tilde{\rho}_{\rm{d}}+1+\tilde{\rho}_{\rm{g}}}\biggr)\Biggr\}, \label{t_settle}
\end{equation}
\\
\noindent
using a rough approximation described in Appendix A. If $\tilde{\rho}_{\rm{d}} \gg 1 + \tilde{\rho}_{\rm{g}}$: the vertical component of the self-gravity of the dust in the dust layer is much larger than the sum of vertical components of the central star's gravity and  the self-gravity of the gas, the dust settling time is approximately equal to $[1-(1/e)]/(\tilde{\rho}_{\rm{d}} \tilde{t}_{\rm{f}})$. On the other hand, if $\tilde{\rho}_{\rm{d}} \ll 1 + \tilde{\rho}_{\rm{g}}$, $\tilde{t}_{\rm{settle}} \approx 1 / [(1 + \tilde{\rho}_{\rm{g}})\tilde{t}_{\rm{f}}]$, and $\tilde{t}_{\rm{settle}}$ results in $1/\tilde{t}_{\rm{f}}$ for the minimum mass solar nebula model because $\tilde{\rho}_{\rm{g}} \ll 1$. Figure \ref{fig3} shows the dust settling time with $\tilde{t}_{\rm{f}} = 0.01$ and $0.1$.

The growth time of the GI is given by the reciprocal of its growth rate. The growth rate is obtained using the method written in \S3 and \S 4 of \citet{ys04} by assuming that the dust density distribution in the direction vertical to the midplane is kept to be Gaussian regardless of the value of the dust density at the midplane. Although the dust density distribution will diverge from the Gaussian as dust aggregates settle, this assumption is fairly good for our rough estimate. The growth time of the GI with the Gaussian dust density distribution is shown in Figure \ref{fig3} for comparison with the dust settling time. After the dust density exceeds the critical density, the growth time of the GI decreases rapidly, and is proportional to the dust density to the power $-1/2$ for $\tilde{\rho}_{\rm{d}} \gg \tilde{\rho}_{\rm{c}}$. On the other hand, the dust settling time is the constant value for $\tilde{\rho}_{\rm{d}} \lesssim \tilde{\rho}_{\rm{g}}$, and is proportional to the dust density to the power $-1$ for $\tilde{\rho}_{\rm{d}} \gg \tilde{\rho}_{\rm{g}}$. Figure \ref{fig3} shows that the GI grows faster than dust settling up to thirty times of the critical density for $\tilde{t}_{\rm{f}} = 0.01$; on the other hand, dust settling is always faster than the growth of the GI for $\tilde{t}_{\rm{f}} = 0.1$. The results of the approximate analytical calculation agree with those of the numerical simulations.

\section{Summary}

We have investigated the density evolution in the dust layer due to the GI and dust settling by numerical simulations and the analytical calculation. If $\tilde{t}_{\rm{f}}=0.01$, the numerical simulation shows that the GI grows faster than dust settling. This path of the GI is considered to be analogous with the case of \citet{sek83} applying the one-fluid model which the dust and gas have the same velocity due to the strong coupling by the gas drag force, although our simulations take into account the velocity difference of the dust and gas. On the other hand, if $\tilde{t}_{\rm{f}}=0.1$, the numerical simulation shows that dust settling is faster than the growth of the GI as long as the dust density at the midplane is smaller than about twenty times of the critical density of the GI. In this case, the approximate analytical calculation reveals that dust settling is faster than the growth of the GI regardless of the dust density at the midplane, so the dust layer will become extremely thin. Subsequently, the GI will grow in the extreme thin dust layer with much smaller wavelengths than the critical wavelength, and then modes with larger wavelengths may begin to grow slowly. This path of the GI is considered to be analogous with the case of \citet{gw73}. Therefore, our results indicate that the difference of the nondimensional gas friction time leads to two evolutional paths of the axisymmetric GI.

In order to confirm whether the planetesimal formation by the GI is possible or not, nonaxisymmetric calculations are essential. We plan to perform the numerical simulations and investigate three dimensional evolution of the GI in subsequent papers.

\acknowledgments

Numerical simulations have been performed on VPP5000 at the Astronomical Data Center of the National Astronomical Observatory of Japan and on eServer p5 595 at Computing and Communications Center of Kyushu University. F. Y. is supported by the Research Fellowship of Japan Society for the Promotion of Science (JSPS) for Young Scientists. M. S. is supported by Grants-in-Aid of the Japanese Ministry of Education, Culture, Sports, Science, and Technology and of JSPS (17039010 and 16340174).

\appendix

\section{DERIVATION OF EQUATION (2)}

Here, we show how to derive equation (\ref{t_settle}). Although the dust density is not constant for $z$, we here assume $\partial \rho_{\rm{d}}/\partial z=0$ and $\partial(\rho_{\rm{d}} u_{\rm{d}})/\partial x=0$ for a rough estimate of the growth time of the dust density. Then the continuity equation of the dust is written

\begin{equation}
  \frac{\partial{\rho}_{\rm{d}}}{\partial{t}}
    +\rho_{\rm{d}}\frac{\partial{w}_{\rm{d}}}{\partial{z}}=0. \label{rhod-App}
\end{equation}
\\
\noindent
Substituting equation (\ref{wd}) into equation (\ref{rhod-App}) and using nondimensional variables, we have

\begin{equation}
  \frac{d \tilde{\rho}_{\rm{d}}}{d\tilde{t}}
    = \tilde{\rho}_{\rm{d}} \tilde{t}_{\rm{f}} (\tilde{\rho}_{\rm{d}}+1+\tilde{\rho}_{\rm{g}}). \label{drhod-App}
\end{equation}
\\
\noindent
Integrating equation (\ref{drhod-App}) with respect to $\tilde{t}$, we have

\begin{equation}
  \tilde{t} = \frac{1}{(1+\tilde{\rho_{\rm{g}}})\tilde{t}_{\rm{f}}}
    \mbox{ln}\biggl(\frac{\tilde{\rho_{\rm{d}}}}{\tilde{\rho_{\rm{d}}}+1+\tilde{\rho_{\rm{g}}}}\biggr)+C, \label{t-App}
\end{equation}
\\
\noindent
where $C$ is an arbitrary constant of integration. From equation (\ref{t-App}), the dust settling time during which the dust density increases by a factor of $e$ results in equation (\ref{t_settle}).


\vspace*{0.2in}

\begin{thebibliography}{}
\bibitem[Blum \& Wurm(2000)]{bw00} Blum, J., \& Wurm, G. 2000, Icarus, 143, 138
\bibitem[Blum et al.(2000)]{bet00} Blum, J., et al. 2000, Phys. Rev. Lett., 85, 2426
\bibitem[Cuzzi, Dobrovolskis, \& Champney(1993)]{cdc93} Cuzzi, J. N., Dobrovolskis, A. R., \&
Champney, J. M. 1993, Icarus, 106, 102
\bibitem[Dobrovolskis, Dacles-Mariani, \& Cuzzi(1999)]{ddc99} Dobrovolskis, A. R.,
Dacles-Mariani, J. S., \& Cuzzi, J. N. 1999, J. Geophys. Res., 104, 30805
\bibitem[Goldreich \& Ward(1973)]{gw73} Goldreich, P., \& Ward, W. R. 1973, \apj, 183, 1051
\bibitem[Haghighipour \& Boss(2003)]{hb03} Haghighipour, N., \& Boss, A. P. 2003, \apj, 583, 996
\bibitem[Hayashi \& Nakagawa(1975)]{hn75} Hayashi, C., \& Nakagawa, Y. 1975, Prog. Theor. Phys., 54, 93
\bibitem[Hayashi(1981)]{hay81} Hayashi, C. 1981, Prog. Theor. Phys. Suppl., 70, 35
\bibitem[Hayashi, Nakazawa, \& Nakagawa(1985)]{hnn85} Hayashi, C., Nakazawa, K., \& Nakagawa, Y. 1985, in Protostars and Planets II, ed. D. C. Black \& M. S. Mathews (Tucson: Univ. Arizona Press), 1100
\bibitem[Nakagawa, Sekiya, \& Hayashi(1986)]{nsh86} Nakagawa, Y., Sekiya, M., \& Hayashi, C. 1986, Icarus, 67, 375
\bibitem[Nomura \& Nakagawa(2006)]{nn06} Nomura, H., \& Nakagawa, Y. 2006, \apj, 640, 1099
\bibitem[Safronov(1969)]{saf69} Safronov, V. S. 1969, Evolution of the Protoplanetary Cloud and Formation of the Earth and Planets (Moscow: Nauka)
\bibitem[Sekiya(1983)]{sek83} Sekiya, M. 1983, Prog. Theor. Phys., 69, 1116
\bibitem[Sekiya(1998)]{sek98} Sekiya, M. 1998, Icarus, 133, 298
\bibitem[Sirono(2004)]{sir04} Sirono, S. 2004, Icarus, 167, 431
\bibitem[Weidenschilling(1980)]{wei80} Weidenschilling, S. J. 1980, Icarus, 44, 172
\bibitem[Weidenschilling(1995)]{wei95} Weidenschilling, S. J. 1995, Icarus, 116, 433
\bibitem[Weidenschilling \& Cuzzi(1993)]{wc93} Weidenschilling, S. J., \& Cuzzi, J. N. 1993, in Protostars and Planets III, ed. E. H. Levy \& J. L. Lunine (Tucson: Univ. Arizona Press), 1031
\bibitem[Yamoto \& Sekiya(2004)]{ys04} Yamoto, F., \& Sekiya, M. 2004, Icarus, 170, 180
\bibitem[Youdin \& Shu(2002)]{ys02} Youdin, A. N., \& Shu, F. H. 2002, \apj, 580, 494
\bibitem[Youdin \& Chiang(2004)]{yc04} Youdin, A. N., \& Chiang, E. I. 2004, \apj, 601, 1109
\end{thebibliography}
\end{document}